\documentclass[aps,pra,preprint,groupedaddress,amsmath,amssymb]{revtex4}

\usepackage{hyperref}

\newcommand{\point}[1]{\ensuremath{{\rm #1}}}
\newcommand{\subP}{_\point{P}}
\newcommand{\ie}{{\em i.e.,\ }}
\newcommand{\eg}{{\em e.g.,\ }}

\newcommand{\Ref}[1]{Ref.~\onlinecite{#1}}

\begin{document}
\title{Morphology of the nonspherically decaying radiation
generated by a rotating superluminal source: reply to comment}

\author{Houshang Ardavan}
\affiliation{Institute of Astronomy, University of Cambridge,
Madingley Road, Cambridge CB3 0HA, UK}
\author{Arzhang Ardavan}
\affiliation{Clarendon Laboratory, Department of Physics, University of Oxford,
Parks Road, Oxford OX1 3PU, UK}
\author{John Singleton}
\affiliation{National High Magnetic Field Laboratory, MS-E536,
Los Alamos National Laboratory, Los Alamos, New Mexico 87545}
\author{Joseph Fasel}
\author{Andrea Schmidt}
\affiliation{Process Modeling and Analysis, MS-F609
Los Alamos National Laboratory, Los Alamos, New Mexico 87545}

\date{2008 April 30}

\begin{abstract}
The fact that the formula used by Hannay in the preceding 
Comment is ``from a standard text on electrodynamics''
neither warrants that it is universally applicable, 
nor that it is unequivocally correct. 
We have explicitly shown [J.\ Opt.\ Soc.\ Am.\ A {\bf 25}, 543 (2008)] 
that,
since it does not include the boundary contribution
toward the value of the field,
the formula in question is not applicable
when the source is extended
and has a distribution pattern that rotates faster than light {\it in vacuo}.
The neglected boundary term
in the retarded solution to the wave equation
governing the electromagnetic field
forms the basis of diffraction theory. 
If this term were identically zero,
for the reasons given by Hannay, 
the diffraction of electromagnetic waves through apertures
on a surface enclosing a source would have been impossible. 
\end{abstract}
\maketitle

\section{Introduction}
The argument presented by Hannay in \Ref{HannayJH:Morphology} 
is based on an incorrect solution of Maxwell's equations.
We have explicitly shown \cite{ArdavanH:Bound} 
that the retarded solution to the wave equation
\begin{equation}
{\bf\nabla}^2{\bf B}-{1\over c^2}{\partial^2{\bf B}\over\partial t^2}=
-{4\pi\over c}{\bf\nabla\times j}.
\label{eq:1}
\end{equation}
governing the magnetic field ${\bf B}$ is {\em not} always given by
\begin{equation}
{\bf B}({\bf x}_P,t_P)=\frac{1}{c}\int{\rm d}^3x
\frac{[{\bf\nabla\times j}]}{\vert{\bf x}_P-{\bf x}\vert},
\label{eq:2}
\end{equation}
as assumed by Hannay \cite{HannayJH:Morphology};
an exception is the solution of Maxwell's equations
describing the emission from a polarization current density ${\bf j}$
whose distribution pattern rotates superluminally
(\ie faster than light {\em in vacuo}). 
[Here,
$({\bf x}_P,t_P)$ and $({\bf x},t)$
are the space-time coordinates of the observation point and the source points, respectively,
$c$ is the speed of light {\em in vacuo},
and the square brackets denote the retarded value of ${\bf\nabla\times j}$.]
The emission from such a source consists of a collection of narrowing subbeams
for which the absolute value of the gradient of the radiation field ${\bf B}$
{\em increases} (as ${R\subP}^{7/2}$)
with the distance $R\subP$ from its source, 
rather than decreasing as predicted by 
Eq.\ (\ref{eq:2}) \cite{ArdavanH:Bound}.  
The inadequacy of Eq.\ (\ref{eq:2}), in this case, lies in 
the neglect of the boundary contribution toward the value of ${\bf B}$.
 
We first describe how the boundary contribution
to the retarded solution for the {\em potential}
can always be made equal to zero,
irrespective of the source motion.
In the Lorenz gauge,
the electromagnetic fields
\begin{equation}
{\bf E}=-\nabla\subP A^0-\partial{\bf A}/\partial(c t\subP),
\qquad{\bf B}=\nabla\subP\times{\bf A},
\label{eq:3}
\end{equation}
are given by a four-potential $A^\mu$
that satisfies the wave equation
\begin{equation}
{\bf\nabla}^2A^\mu-{1\over c^2}{\partial^2A^\mu\over\partial t^2}=
-{4\pi\over c}j^\mu,\qquad\mu=0,\cdots, 3,
\label{eq:4}
\end{equation}
where $A^0/c$ and $j^0/c$ are the electric potential and charge density
and $A^\mu$ and $j^\mu$ for $\mu=1,2,3$
are the Cartesian components of the magnetic potential ${\bf A}$
and the current density ${\bf j}$ \cite{JacksonJD:Classical}.
The solution to the initial-boundary value problem for Eq.\ (\ref{eq:4})
is given by
\begin{equation}\begin{split}
A^\mu({\bf x}\subP,t\subP)=
&{1\over c}\int_0^{t\subP}{\rm d}t\int_V{\rm d}^3x\,j^\mu G
+{1\over4\pi}\int_0^{t\subP}{\rm d}t
\int_\Sigma{\rm d}{\bf S}\cdot(G\nabla A^\mu-A^\mu\nabla G)\\
&-{1\over 4\pi c^2}\int_V{\rm d}^3x
\Big(A^\mu{\partial G\over\partial t}-G{\partial A^\mu\over\partial t}\Big)_{t=0},
\end{split}\label{eq:5}
\end{equation}
in which $G$ is the Green's function
and $\Sigma$ is the surface enclosing the volume $V$
(see, \eg page 893 of \Ref{MorsePM:Methods1}).

The potential arising from a general time-dependent localized source in unbounded space
decays as ${R\subP}^{-1}$ when $R\subP\equiv\vert{\bf x}\subP\vert\to\infty$,
so that for an arbitrary free-space potential
the second term in Eq.\ (\ref{eq:5})
would be of the same order of magnitude
($\sim{R\subP}^{-1}$)
as the first term,
in the limit that the boundary $\Sigma$ tends to infinity.
However,
even potentials that satisfy the Lorenz condition
${\bf\nabla\cdot A}+c^{-2}\partial A^0/\partial t=0$
are arbitrary
to within a solution of the homogeneous wave equation:
the gauge transformation
\begin{equation}
{\bf A}\to{\bf A}+\nabla\Lambda,
\qquad A^0\to A^0-\partial\Lambda/\partial t
\label{eq:6}
\end{equation}
preserves the Lorenz condition
if $\nabla^2\Lambda-c^{-2}\partial^2\Lambda/\partial t^2=0$
(see \Ref{JacksonJD:Classical}).
One can always use this gauge freedom in the choice of the potential
to render the boundary contribution
(the second term)
in Eq.\ (\ref{eq:5})
equal to zero,
since this term, too,
satisfies the homogeneous wave equation.
Under the null initial conditions
$A^\mu|_{t=0}=(\partial A^\mu/\partial t)_{t=0}=0$
assumed in this note,
the contribution from the third term in Eq.\ (\ref{eq:5})
is identically zero.

In the absence of boundaries,
the retarded Green's function has the form
\begin{equation}
G({\bf x}, t;{\bf x}\subP, t\subP)={\delta(t\subP-t-R/c)\over R},
\label{eq:7}
\end{equation}
where $\delta$ is the Dirac delta function
and $R$ is the magnitude of the separation
${\bf R}\equiv{\bf x}\subP-{\bf x}$
between the observation point ${\bf x}\subP$
and the source point ${\bf x}$.
Irrespective of whether the radiation decays spherically 
(as in the case of a conventional source) or nonspherically
(as would apply for a rotating superluminal source) \cite{ArdavanH:Bound},
therefore,
the potential $A^\mu$ due to a localized source distribution
that is switched on at $t=0$ in an unbounded space,
can be calculated from the first term in Eq.\ (\ref{eq:5}):
\begin{equation}
A^\mu({\bf x}\subP,t\subP)=
c^{-1}\int{\rm d}^3 x{\rm d}t\, j^\mu({\bf x},t)\delta(t\subP-t-R/c)/R,
\label{eq:8}
\end{equation}
\ie from the classical expression for the retarded potential.
Whatever the Green's function for the problem may be
in the presence of boundaries,
it would approach that in Eq.\ (\ref{eq:7})
in the limit where the boundaries tend to infinity,
so that one can also use this potential to calculate the field
on a boundary that lies at large distances from the source.

We now turn to the case of the {\em field} and show that an analogous
assumption about the boundary contribution may not be made. 
Consider the wave equation (\ref{eq:1}) governing the magnetic field;
Equation (\ref{eq:1}) may be obtained
by simply taking the curl of the wave equation for the vector potential
[Eq.\ (\ref{eq:4}) for $\mu=1,2,3$].
We write the solution to the initial-boundary value problem for Eq.\ (\ref{eq:1}),
in analogy with Eq.\ (\ref{eq:5}),
as
\begin{equation}\begin{split}
B_k({\bf x}\subP,t\subP)=
&{1\over c}\int_0^{t\subP}{\rm d}t\int_V{\rm d}^3x\,({\bf\nabla\times j})_k G
+{1\over4\pi}\int_0^{t\subP}
{\rm d}t\int_\Sigma{\rm d}{\bf S}\cdot(G\nabla B_k-B_k\nabla G)\\
&-{1\over 4\pi c^2}\int_V{\rm d}^3x
\Big(
B_k{\partial G\over\partial t}-G{\partial B_k\over\partial t}
\Big)_{t=0},
\end{split}\label{eq:9}
\end{equation}
where $k=1,2,3$ designate the Cartesian components of ${\bf B}$ and ${\bf\nabla\times j}$.

Here, we no longer have the freedom,
offered by a gauge transformation in the case of Eq.\ (\ref{eq:5}),
to make the boundary term zero. 
Nor does this term always decay faster than the source term,
so that it could be neglected for a boundary that tends to infinity,
as is commonly assumed in textbooks
(\eg page 246 of \Ref{JacksonJD:Classical})    
and the published literature \cite{HannayJH:Bouffr,HannayJH:ComIGf,HannayJH:ComMhd,HannayJH:Speapc}.
The boundary contribution
to the retarded solution of the wave equation governing the field
[the second term on the right-hand side of Eq.\ (\ref{eq:9})]
entails a surface integral
over the boundary values of both the field and its gradient.
For a rotating superluminal source,
where the gradient of the field increases
as ${R\subP}^{7/2}$
over a solid angle that decreases as ${R\subP}^{-4}$,
this boundary contribution
turns out to be proportional to ${R\subP}^{-1/2}$
(see \Ref{ArdavanH:Bound}).
Not only is this not negligible
relative to the contribution from the source term
[the first term on the right-hand side of Eq.\ (\ref{eq:9})],
but the boundary term constitutes the dominant contribution
toward the value of the radiation field in this case.

If one ignores the boundary term
in the retarded solution of the wave equation governing the field
(as Hannay does \cite{HannayJH:Bouffr,HannayJH:ComIGf,HannayJH:ComMhd,HannayJH:Speapc,HannayJH:Morphology}),
one obtains a different result,
in the superluminal regime,
from that obtained
by calculating the field via the retarded potential \cite{ArdavanH:Genfnd,ArdavanH:Speapc,ArdavanH:Morph}.
This apparent contradiction
stems solely from having ignored a term
in the solution to the wave equation
that is by a factor of the order of ${R\subP}^{1/2}$ greater
than the term that is normally kept in this solution.
The contradiction disappears
once the neglected term is taken into account:
the solutions to the wave equations
governing both the potential and the field
predict that the field of a rotating superluminal source
decays as ${R\subP}^{-1/2}$
as $R\subP$ tends to infinity, a result that has now been demonstrated experimentally~
\cite{ArdavanA:Exponr}.

For a volume $V$ in which there are no sources,
Eq.\ (\ref{eq:9}) reduces to 
\begin{equation}
B_k({\bf x}\subP,t\subP)=
{1\over4\pi}\int_0^{t\subP}
{\rm d}t\int_\Sigma{\rm d}{\bf S}\cdot(G\nabla B_k-B_k\nabla G),
\label{eq:10}
\end{equation}
under the null initial conditions $B_k|_{t=0}=(\partial B_k/\partial t)_{t=0}=0$.
As in the customary geometry for diffraction,
the closed surface $\Sigma$ can consist of two disjoint closed surfaces,
$\Sigma_{\rm inner}$ and $\Sigma_{\rm outer}$
(\eg two concentric spheres),
both of which enclose the source
(see Fig.~10.7 of \Ref{JacksonJD:Classical}).
If the observation point does not lie in the region between $\Sigma_{\rm inner}$ and $\Sigma_{\rm outer}$,
\ie lies outside the closed surface $\Sigma$,
then the composite surface integral in Eq.\ (\ref{eq:10}) vanishes:   
\begin{equation}\begin{split}
\int_0^{t\subP}
{\rm d}t\int_\Sigma{\rm d}{\bf S}\cdot(G\nabla B_k-B_k\nabla G)
&=\int_0^{t\subP}
{\rm d}t\left(\int_{\Sigma_{\rm inner}}+\int_{\Sigma_{\rm outer}}\right){\rm d}{\bf S}\cdot(G\nabla B_k-B_k\nabla G)\\
&=0
\end{split}\label{eq:11}
\end{equation}
(see \Ref{MorsePM:Methods1}).
But under no circumstances
would the integrals over $\Sigma_{\rm inner}$ or $\Sigma_{\rm outer}$ vanish individually
if these surfaces enclose a source,
\ie if there is a nonzero field inside $\Sigma_{\rm inner}$.
Nor does the fact that the values of these integrals are unchanged
by deformations of $\Sigma_{\rm inner}$ and $\Sigma_{\rm outer}$
have any bearing on whether they are nonvanishing or not.
Equation (\ref{eq:10}) forms the basis of diffraction theory \cite{JacksonJD:Classical}.
If the surface integrals over $\Sigma_{\rm inner}$ and $\Sigma_{\rm outer}$
were to vanish individually,
as claimed by Hannay \cite{HannayJH:Morphology},
the diffraction of electromagnetic waves
through apertures on a surface enclosing a source
would have been impossible
(see Sec.~10.5 of \Ref{JacksonJD:Classical}).

Finally,
we must stress that what one obtains by including the boundary term
in the retarded solution to the wave equation governing the field
is merely a mathematical identity;
it is not a solution
that could be used to calculate the field
arising from a given source distribution in free space.
Unless its boundary term
happens to be small enough relative to its source term to be neglected,
a condition that cannot be known {\em a priori},
the solution in question would require
that one prescribe the field in the radiation zone
(\ie what one is seeking)
as a boundary condition.
Thus,
the role played by the classical expression
for the retarded potential in radiation theory
is much more fundamental
than that played by the corresponding retarded solution
of the wave equation governing the field.
The only way to calculate the free-space radiation field
of an accelerated superluminal source
is to calculate the retarded potential
and differentiate the resulting expression
to find the field (see also \Ref{ArdavanH:Speapc1}).

\section*{Acknowledgements}
A.\ A.\ is supported by the Royal Society.
J.\ S., J.\ F., and A.\ S.\ are supported by U.S.\ Department of Energy grant LDRD 20080085DR,
``Construction and use of superluminal emission technology demonstrators
with applications in radar, astrophysics and secure communications.''

\bibliography{jabosa,superluminal}

\end{document}